\documentclass[twoside]{article}
\usepackage{fleqn,espcrc2}
\usepackage{epsfig,epsf}

\usepackage{graphicx}

\newcommand{\bee}{\begin{equation}}
\newcommand{\ee}{\end{equation}}

\newcommand{\AmS}{{\protect\the\textfont2
  A\kern-.1667em\lower.5ex\hbox{M}\kern-.125emS}}

\title{Nontrivial fixed point in nonabelian models}

\author{Adrian Patrascioiu
           \address {Physics Department, University of Arizona,\\
        Tucson, AZ 85721, U.S.A.}
  and
   Erhard Seiler\address{Max-Planck-Institut f\"ur Physik,
  Werner-Heisenberg-Institut,\\
        F\"ohringer Ring 6, 80805 Munich, Germany}
        \thanks{Speaker}
}
\begin{document}

\begin{abstract}
We investigate the percolation properties of equatorial strips in the
two-dimensional O(3) nonlinear $\sigma$ model. We find convincing
evidence that such strips do not percolate at low temperatures,
provided they are sufficiently narrow. Rigorous arguments show that
this implies the vanishing of the mass gap at low temperature and
the absence of asymptotic freedom in the massive continuum limit.
We also give an intuitive explanation of the transition to a
massless phase and, based on it, an estimate of the transition
temperature.

\vspace{1pc}
\end{abstract}

\maketitle

\section{Introduction}

This talk, though scheduled in the session on Perturbation Theory (PT),
does not deal with PT as such. But one of its conclusions is that in 
certain nonabelian models the perturbative Callan-Symanzik 
$\beta$-function gives the wrong picture of the renormalization group 
flow.

In 1991 we developed a rigorous criterion for the existence of a 
massless phase in $2D$ spin models based on percolation properties 
of certain subsets of the target spin manifold \cite{psperc}. This was
presented at the 1992 lattice conference together with nonrigorous 
arguments that led to the conclusion that all $2D$ $O(N)$ most 
likely had a soft low temperature phase, contrary to prevailing 
expectations.

Advances in computer technology have made it feasible to tackle
directly the question whether the percolation properties leading
to the existence of a massless phase hold or not. In this talk
I report our results giving direct numerical evidence for our old 
conjecture of the existence of a massless phase in models for which 
PT predicts mass generation and asymptotic freedom (AF); these results
were first presented in \cite{psabs}).

For concretenss and simplicity we are dealing with the $2D$ $O(3)$
model, a.k.a. classical Heisenberg model. Its standard version
is defined on the square lattice, with the standard Hamiltonian
(action)
\bee
H=-\sum_{\langle xy\rangle} s(x)\cdot s(y)
\ee
where $s(.)$ is a classical spin (unit 3-vector) living on the
vertices $x$ of the lattice and the sum is over nearest neighbors
${\langle xy\rangle}$; the spins are distributed according to the
Gibbs measure with density ${1\over Z}\exp(-\beta H)$.
For the purpose of this investigation we modify this model by 
introducing a constraint limiting the size of the deviation
allowed between neighboring spins:
\bee
s(x)\cdot s(y)\geq c \label{constraint}
\ee
with $-1\leq c\leq 1$ (the so-called cut action with cut $c$), and 
we also replace the square by a triangular lattice; the Gibbs measure
thus contains $\theta$-functions enforcing the constraint 
(\ref{constraint}).
These modifications are made for purely technical reasons and we have
checked that they do not change the universality class \cite{psprep}.

\section{The percolation criterion}
We briefly sketch the percolation criterion developed in 1991
\cite{psperc}:
We divide the sphere $S_2$ into three pieces:
\begin{itemize}
\item
`equatorial strip' ${\cal S}_\epsilon$, defined by
$|s\cdot n|<\epsilon/2$ for some fixed unit vector $n$.
\item
`upper polar cap' ${\cal P}_\epsilon^+$, defined by
$s\cdot n\geq\epsilon/2$,
\item
`lower polar cap' ${\cal P}_\epsilon^-$, defined by
$s\cdot n\leq -\epsilon/2$.
\end{itemize}
and denote the corresponding subsets of the lattice for a given
configuration by $S_\epsilon^\pm$ etc.. These subsets
fall into connected components called `clusters'; their mean size 
we denote by $\langle S_\epsilon\rangle$ etc. If this mean size
is finite, we say that the clusters `form islands'; if there
is an infinite cluster, we say the subset percolates.
There is a third possibility: that the clusters have divergent mean
size, but none of them is infinite; this we call `formation of rings'.

The main result of \cite{psperc} is the following

{\it Theorem:} If for a certain $c>1-\epsilon^2/2$ \ \
$S_\epsilon$ does not percolate, the model has no mass gap.

The reason why this is so can be understood by using the
Fortuin-Kasteleyn (FK) representation \cite{FK} for the imbedded
Ising spins $\sigma_x\equiv sgn(s(x))$ (which is also the basis of 
the cluster algorithm).

But first one has to notice that in $2D$ it is not possible that
two disjoint clusters both percolate, and therefore, if 
$c>1-\epsilon^2/2$, the union of the polar caps cannot
percolate, because then both of them would percolate.
But if also $S_\epsilon$ does not percolate, as assumed,
a lemma of Russo \cite{russo} assures us that the clusters of
each of the three sets have divergent mean size.

But due to the inequality $c>1-\epsilon^2/2$, each of the
clusters of $P_\epsilon^+$  or $P_\epsilon^-$ has
to be contained in its entirety inside a FK cluster, and hance
also the FK clusters have divergent mean size.

Since the mean size of the FK clusters is equal to the susceptibility
of the imbedded Ising spins, this divergent mean size is incompatible
with exponential clustering, and thus there is no mass gap.

\section{Percolation properties: numerical study}
We investigated numerically the percolation properties of
the equatorial strip for the special case $\beta=0$. In this
case the parameter $c$ replaces the temperature in determining
how ordered or disordered the system is.

\begin{figure}[htb]
\centerline{\epsfxsize=6cm\epsfbox{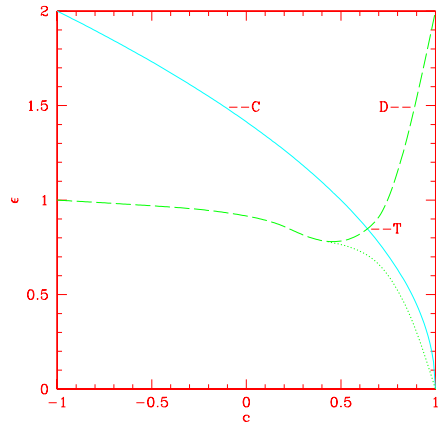}}
\vglue-1cm
\caption{Phase diagram of the $O(3)$ model on the T lattice}
\label{phase}
\end{figure}

The results of our investigation \cite{psabs} are summarized in the 
`percolation phase diagram' Fig.\ref{phase}. The diagram is 
semiquantitative, but qualitatively correct, as we will explain.

In this figure the solid line is the curve $c=1-\epsilon^2/2$;
above that line the two polar caps cannot `touch' and therefore
their union cannot percolate. The dashed line
separates a regime in which the strip $S_\epsilon$
percolates (above) and one in which it does not (below). For small
$c$, the strip forms islands for $\epsilon$ below that line.
Around $c=0.4$ a dotted line branches off; below it the strip still 
forms islands, whereas between the dotted and the dashed lines the
clusters of the strip have mean infinite size without percolating
(`formation of rings'). The interesting region is the one between 
the solid and the dashed lines: here the strip does not percolate 
but the inequality of the theorem holds. So in this regime our theorem 
can be applied and allows us to conclude that for $c>c_o$, 
$c_o\approx 0.7$ there is no mass gap.

\begin{figure}[htb]
\centerline{\epsfxsize=6cm\epsfbox{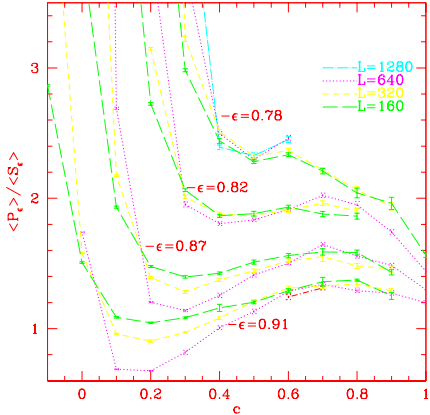}}
\vglue-1cm
\caption{Ratio $\langle P_\epsilon\rangle/\langle S_\epsilon\rangle$
for various $\epsilon$ values versus $c$}
\label{tr}
\end{figure}

Fig.\ref{phase} also shows that for $\epsilon<\epsilon_o$, with
$\epsilon_o\approx 0.76$ the equatorial strip does not percolate for
any $c$. [All\`es et al \cite{alles} published a study
showing that for $\epsilon=1.05$ and $\beta=2.0$ (standard
action) the equatorial strip percolates. This is correct,
but since their choice of parameters is such that they are both
in the massive phase and in the percolation region of the strip,
it is not very relevant for our problem].

Let me now explain from which facts our `phase diagram' was derived:
In Fig.\ref{tr} we show that ratios
$r\equiv\langle P_\epsilon\rangle/\langle S_\epsilon\rangle$
as a function of $c$ for several values of $\epsilon$ between .78 
and .91 and for lattice sizes $L$ from 160 to 1280.

\begin{figure}[htb]
\centerline{\epsfxsize=6cm\epsfbox{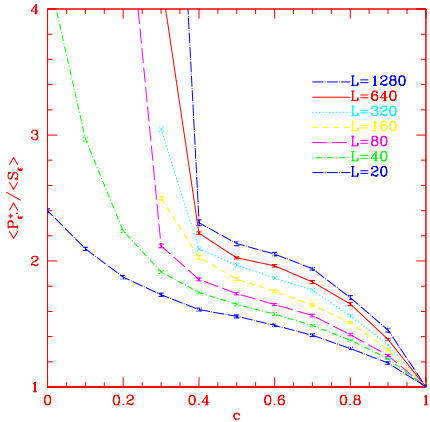}}
\vglue-1cm
\caption{Ratio $\langle P_{0.5}\rangle/\langle S_{0.75}\rangle$}
\label{rings}
\end{figure}

It is seen that for small $c$ (depending on $\epsilon$) $r$ increases 
sharply with $L$ (we have data which are off the scale of this figure 
and show that the increase continues). This expresses the fact that 
in this regime $S_\epsilon$ percolates, wheras its complement 
$P_\epsilon$ forms islands of finite size. At a certain value of $c$ 
the curves for different $L$ intersect and $r$ becomes scale invariant; 
this is the critical point of percolation for the chosen value of 
$\epsilon$ in which both sets form rings. The pair $(c,\epsilon)$ 
defines a point on the dashed curve in Fig.\ref{phase}.

If we increase $c$ beyond the intersection point, the size 
dependence of $r$ is reversed, indicating that we are now in the
regime of percolation of $P_\epsilon$.

\begin{figure}[htb]
\centerline{\epsfxsize=6cm\epsfbox{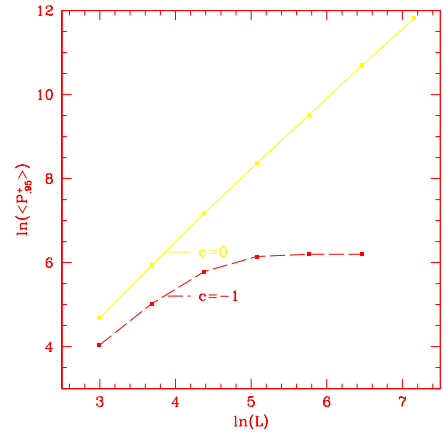}}
\vglue-1cm
\caption{$\ln\langle P_{0.95}\rangle$ vs $\ln L$ for $c=0$ and
$c=-1$}
\label{power}
\end{figure}

Increasing $c$ still further, the curves come together again and
remain together, indicating that for all $c$ in that regime
(depending on $\epsilon$) we are in the regime of rings formation
of both sets, i.e. we have entwred the regime between the dashed 
and dotted curves in Fig. \ref{phase}. For $c\to 1$ the ratio $r$ 
converges to the geometric ratio $(2-\epsilon)/\epsilon$ of the sets 
${\cal P}_\epsilon$ and ${\cal S}_\epsilon$.

To corroborate that for $\epsilon$ less than about 0.76 $S_\epsilon$
does not percolate, no matter what $c$ is, we also measured the ratio
$r'\equiv \langle P_{\epsilon'}^+\rangle/\langle S_\epsilon \rangle$,
where $\epsilon'$ is chosen in such a way that the two sets have
equal area, i.e. $\epsilon'=2-2\epsilon=0.5$. Fig.\ref{rings}
shows that
for $c<0.4$ $r'$ increases sharply with $L$. Since the polar cap 
cannot percolate, this means that $P_\epsilon'$ forms rings of 
arbitrary size, whereas $S_\epsilon$ forms islands of finite size. 
The behavior changes drastically around $c=0.4$, the $L$ dependence of 
the ratio $r'$ becomes a much milder increase, compatible with a 
power law behavior of both numerator and denominator, but clearly 
ruling out percolation of $S_\epsilon$. The only possible
interpretation is that both sets form rings of arbitrary size.

The crucial fact for our conclusion is obviously that a polar cap
can form rings of arbitrary size even though it is smaller than
a hemisphere. This ring formation then prevents percolation of the
corresponding equatorial strip, and thus allows the application
of our theorem. 

Our approach has been by necessity in the spirit of finite size 
scaling, studying how various
quantities change with increasing size. So is it conceivable that
we are deceived by finite size behavior that changes its character
at some astronomical lattice size? Obviously in the truly interesting
region near the critical point (which is around or slightly below
$c=0.7$) we cannot work on lattices of thermodynamic size. But we did 
an additional test at $c=0$, where the correlation length is about 53
and we can easily go to thermodynamic lattices: We measured directly 
the mean cluster size $\langle P_{0.95}$ as a function of $L$ for $L$ 
up to 1280. The results displayed in Fig.\ref{power} show a linear
dependence of $\ln \langle P_{0.95}\rangle $ on $\ln(L)$, indicating a
powerlike increase of $\langle P_{0.95}\rangle$ for lattices much larger
than the correlation length. This is in sharp contrast with the
behavior at $c=-1$, also shown in Fig.\ref{power}, where one can 
clearly see $\langle P_{0.95}\rangle$ leveling off.

If we combine this with another fact, which is plausible and which 
we also checked, namely that $\langle P_\epsilon \rangle$ is a
monotonically increasing function of $c$, we reach the conclusion that
this ring formation must persist for all $c<1$.

\section{Concluding remarks}

Our main conclusion is that the $2D$ $O(3)$ has a transition to
a massless phase at low temperature, contrary to standard lore,
which derives from the PT calculation of the Callan-Symanzik
$\beta$-function. We have questioned the validity of PT in the
models showing perturbative asymptotic freedom in various publications
and talks at lattice conferences (see for instance \cite{pssi}),
but our percolation study makes the conclusion unavoidable that PT
does not give the correct asymptotic expansion for the $\beta$-function.

A question that has been asked is what is `driving the transition' to 
the massless phase. This ia also answered in our recent paper
\cite{psabs}: it can be understood as the transition froma gas of 
instantons to a gas (or liquid) of super-instantons, which are
the dominant excitations at low temperature and create the disorder 
required by the Mermin-Wagner theorem. In \cite{psabs} we used a
simple-minded energy-entropy argument to give a rough estimate of
the transition temperature for the standard action model; the result
($\beta_{crt}\approx \pi$), is consistent with knwon numerical results.

\end{document}